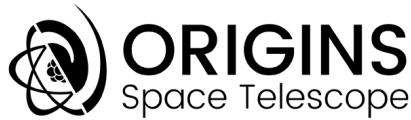

**The Origins Space Telescope: From First Stars to Life**
*The Origins Space Telescope, one of four large Mission Concept studies sponsored by NASA for review in the 2020 US Astrophysics Decadal Survey, will open unprecedented discovery space in the infrared, unveiling our cosmic origins.*

**Authors:** Cara Battersby[1], Lee Armus[2], Edwin Bergin[3], Tiffany Kataria[4], Margaret Meixner[5,6], Alexandra Pope[7], Kevin B. Stevenson[5], Asantha Cooray[8], David Leisawitz[9], Douglas Scott[10], James Bauer[11], C. Matt Bradford[4,12], Kimberly Ennico[13], Jonathan J. Fortney[14], Lisa Kaltenegger[15], Gary J. Melnick[16], Stefanie N. Milam[9], Desika Narayanan[17], Deborah Padgett[4], Klaus Pontoppidan[5], Thomas Roellig[13], Karin Sandstrom[18], Kate Y. L. Su[19], Joaquin Vieira[20], Edward Wright[21], Jonas Zmuidzinas[12], Johannes Staguhn[6,9], Kartik Sheth[22], Dominic Benford[22], Eric E. Mamajek[4], Susan G. Neff[9], Sean Carey[2], Denis Burgarella[23], Elvire De Beck[24], Maryvonne Gerin[25], Frank P. Helmich[26], S. Harvey Moseley[9], Itsuki Sakon[27] and Martina C. Wiedner[25]

[1]Department of Physics, University of Connecticut, Storrs, CT, USA. [2]Infrared Processing and Analysis Center, California Institute of Technology, Pasadena, CA, USA. [3]Department of Astronomy, University of Michigan, Ann Arbor, MI, USA. [4]NASA Jet Propulsion Laboratory, Pasadena, CA, USA. [5]Space Telescope Science Institute, Baltimore, MD, USA. [6]Department of Physics & Astronomy, Johns Hopkins University, Baltimore, MD, USA. [7]Department of Astronomy, University of Massachusetts, Amherst, MA, USA. [8]Department of Physics & Astronomy, University of California, Irvine, CA, USA. [9]NASA Goddard Space Flight Center, Greenbelt, MD, USA. [10]Department of Physics & Astronomy, University of British Columbia, Vancouver, BC, Canada. [11]Department of Astronomy, University of Maryland, College Park, MD, USA. [12]Division of Physics, Mathematics and Astronomy, California Institute of Technology, Pasadena, CA, USA. [13]NASA Ames Research Center, Moffett Field, CA, USA. [14]Department of Astronomy & Astrophysics, University of California, Santa Cruz, CA, USA. [15]Carl Sagan Institute at Cornell, Ithaca, NY, USA. [16]Harvard-Smithsonian Center for Astrophysics, Cambridge, MA, USA. [17]Department of Astronomy, University of Florida, Gainesville, FL, USA. [18]Department of Physics, University of California, San Diego, CA, USA. [19]Steward Observatory, University of Arizona, Tucson, AZ, USA. [20]Department of Astronomy, University of Illinois, Urbana-Champaign, IL, USA. [21]Department of Physics and Astronomy, University of California, Los Angeles, CA, USA. [22]Astrophysics Division, Science Mission Directorate, NASA Headquarters, Washington, DC, USA. [23]Laboratoire d'Astrophysique, University of Marseille, Marseille, France. [24]Department of Space, Earth and Environment, Chalmers Institute of Technology, Onsala Space Observatory, Onsala, Sweden. [25]Sorbonne Université, Observatoire de Paris, Université PSL, CNRS, Paris, France. [26]SRON Netherlands Institute for Space Research, Groningen, The Netherlands. [27]Department of Astronomy, University of Tokyo, Tokyo, Japan.

The Universe has never been seen like this before. The window into the far-infrared opens only above Earth's atmosphere, and humanity has barely glimpsed outside. About half of the light emitted by stars, planets, and galaxies over the lifetime of the Universe emerges in the infrared. With an unparalleled sensitivity increase of up to a factor of 1,000 more than any previous or planned mission, the jump forward offered by the Origins Space Telescope (OST) is akin to that from the naked eye to humanity's first telescope, and the jump from Galileo's first telescope to humanity's first telescope in space. While key path-finding missions have glimpsed a rich infrared cosmos, extraordinary discovery space awaits; the time for a far-IR revolution has begun.

Are we alone or is life common in the Universe? OST directly addresses this long-standing question by searching for signs of life in the atmospheres of potentially habitable terrestrial planets transiting M dwarfs. How do planets become habitable? OST traces the trail of cold water from the interstellar medium, through protoplanetary disks, and into the outer reaches of our own solar system. How do stars, galaxies, black holes, and the elements of life form from cosmic dawn to today? With broad wavelength coverage and fast mapping speeds, OST will map millions of galaxies, simultaneously measuring star formation rates and black hole growth across cosmic time, peering deeper into the far reaches of the Universe than ever before.

OST will be maintained at a temperature of 4 K, enabling its tremendous sensitivity gain, and will operate from 5–600 μm, encompassing the mid- and far-infrared. OST has two Mission Concepts: Concept 1 with a 9.1-m deployed off-axis primary, and Concept 2, described here, a non-deployed 5.9-m on-axis telescope with the equivalent collecting area of the James Webb Space Telescope. Concept 2 includes four instruments with capabilities for imaging (large surveys and pointed), spectroscopy (survey and high-resolution modes), and polarimetry, as well as an instrument for high-precision transiting exoplanet spectroscopy. Concept 2 is optimized for minimal complexity and fast mapping. We describe here the three key science themes for OST and the basic mission specifications.

*Are we alone in the Universe?*

For the first time in human history, our generation will have the technology to answer this age-old question. Planet-hunting programs (such as TRAPPIST [Gillon et al. 2016], MEarth [Berta-Thompson et al. 2013], Kepler [Borucki et al. 2010]) recently confirmed the first Earth analogues orbiting M dwarfs. However, it is unknown whether planets orbiting M dwarfs, the most ubiquitous stars in our Galaxy, can support life. OST will characterize the atmospheres of Earth-size planets, and search for biosignatures, a combination of molecules that, when observed together, can only be sustained by life.

Exoplanets are incredibly dim compared to their host stars. The transit technique is currently the only method that has characterized the atmospheres of potentially habitable worlds (de Wit et al., 2018). During primary transit, we observe the planet pass in front of its host star, which unambiguously determines the planet's size and, when combined with a mass measurement, its bulk density. Knowledge of these fundamental parameters, in addition to temperature, is an essential first step towards assessing its ability to support life. OST will expand upon the legacy of exoplanet science with the Hubble Space Telescope, Spitzer Space Telescope and the upcoming James Webb Space Telescope by using the transit technique to study the climates of Earth-size worlds orbiting within the habitable zones of M-dwarf stars.

The true indicator of life, however, is the detection of biosignatures, a combination of mutually reactive molecules whose presence together in an atmosphere can only be sustained by life. For example, the measurable presence of methane (a reducing molecule) in Earth's oxidizing atmosphere (containing oxygen and ozone) is widely considered a biosignature (Kaltenegger, 2017). OST will include an instrument designed to detect light absorbed and re-radiated by potentially habitable planets, whose emission peaks in the mid-infrared (see Figure 1). By obtaining emission and transmission spectra from 5 to 25 μm, OST will be able to precisely measure a planet's atmospheric thermal structure and search for habitability indicators such as water, carbon dioxide, ozone, and methane. The definitive detection of both ozone and methane in the atmospheres of one or more planets orbiting nearby M dwarfs will be the strongest evidence that we are not alone and, in fact, that life is common in our Galaxy.

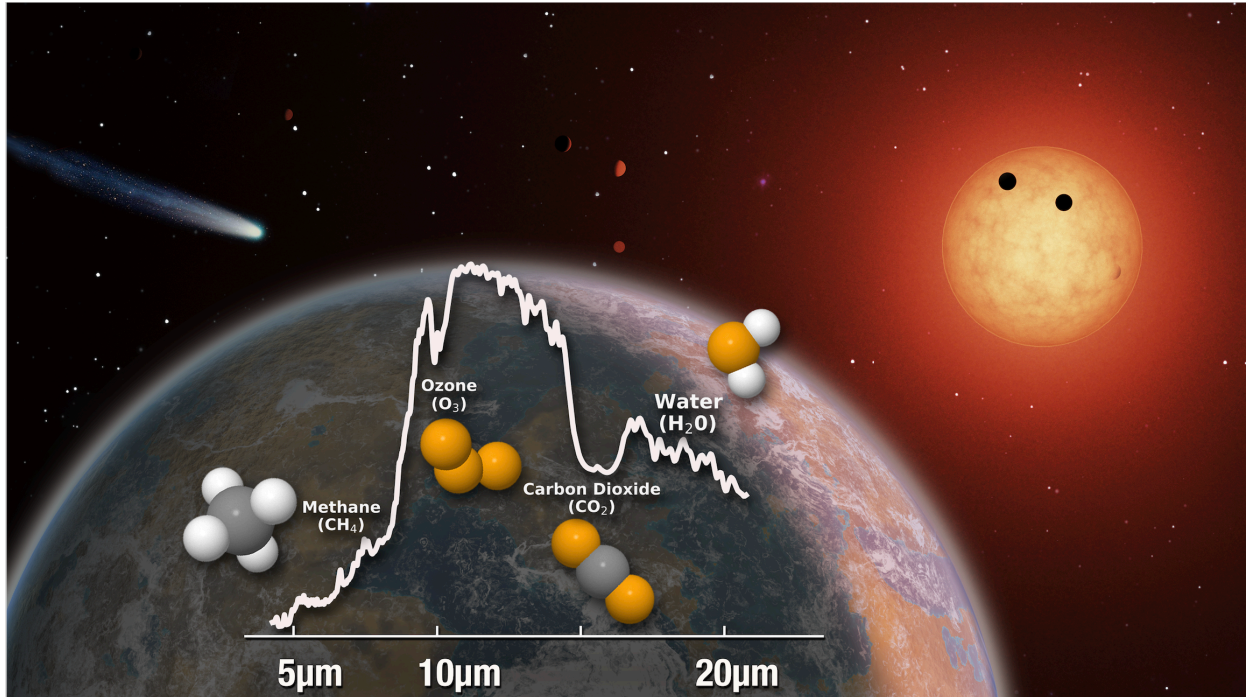

*Figure 1:* *The Origins Space Telescope will search for biosignatures in the atmospheres of potentially habitable terrestrial planets transiting M dwarfs to determine if we are alone in the Universe.*

### *How do planets become habitable?*

The trail of water leads to a habitable world, but it begins in the cold expanse of interstellar space, where water forms in condensing, starless cores. As these starless cores collapse into protoplanetary disks, this water is distributed through the disk, with a transition from a hot (> 200 K) vapor-dominated phase to a cold, ice-dominated phase at the water snowline. While pioneering observations with previous IR facilities (e.g. van Dishoeck et al. 2014) have provided glimpses of this water trail, a complete picture has so far remained elusive.

Uncovering the trail of water requires a space-based platform, such as OST, operating from the mid-infrared, where hundreds of water transitions probe gas at high temperature (T > 400 K), to the far-infrared, where ground-state transitions probe cold (10 < T < 400 K) water vapor. These ground-state transitions are needed to trace the snowline location, which is posited as a favored site of planet formation (Stevenson & Lunine 1988). Within our solar system, OST will survey nearly a hundred comets and provide a true sample of the D/H ratio, and its variations, in cometary reservoirs – key to understanding the origin of water in Earth's oceans.

OST's high sensitivity and broad wavelength coverage will allow us to probe the envelopes of collapsing gas clouds to trace the heretofore hidden supply of water to a natal disk, and simultaneously detect water vapor in hot and cold gas in protoplanetary disks out to the distance of Orion. Along with the ability to probe HD, a tracer of the total $H_2$ mass (Bergin et al. 2013), astronomers will sample the evolving water content, both gas and ice, along with the water snowline as a function of stellar mass and evolutionary state. Ultimately, OST will provide the definitive picture of how water forms and is provided to planetary systems, which, with this central ingredient and others, can lead to life's formation on alien worlds.

*How do stars, galaxies, black holes, and the elements of life form from cosmic dawn to today?*

The Milky Way has spent the last 12 billion years growing into the familiar spiral Galaxy we live in today: its central supermassive black hole growing along with the stars, mostly hidden behind dust (Maddau & Dickinson 2014). In order to understand the complicated relationship between the growth of stars and black holes in galaxies, and chart this evolution over cosmic time, OST will characterize spectroscopically large samples of distant galaxies at the wavelengths where they emit most of their energy, thereby viewing the Universe at earlier and earlier epochs, effectively rewinding the clock on the lifecycle of galaxies. As OST pushes to the earliest epochs, we will observe the rise of metals and dust - the creation and dispersal of heavy elements inside and outside galaxies through successive generations of stellar birth and death - from the first stars and galaxies to the present epoch.

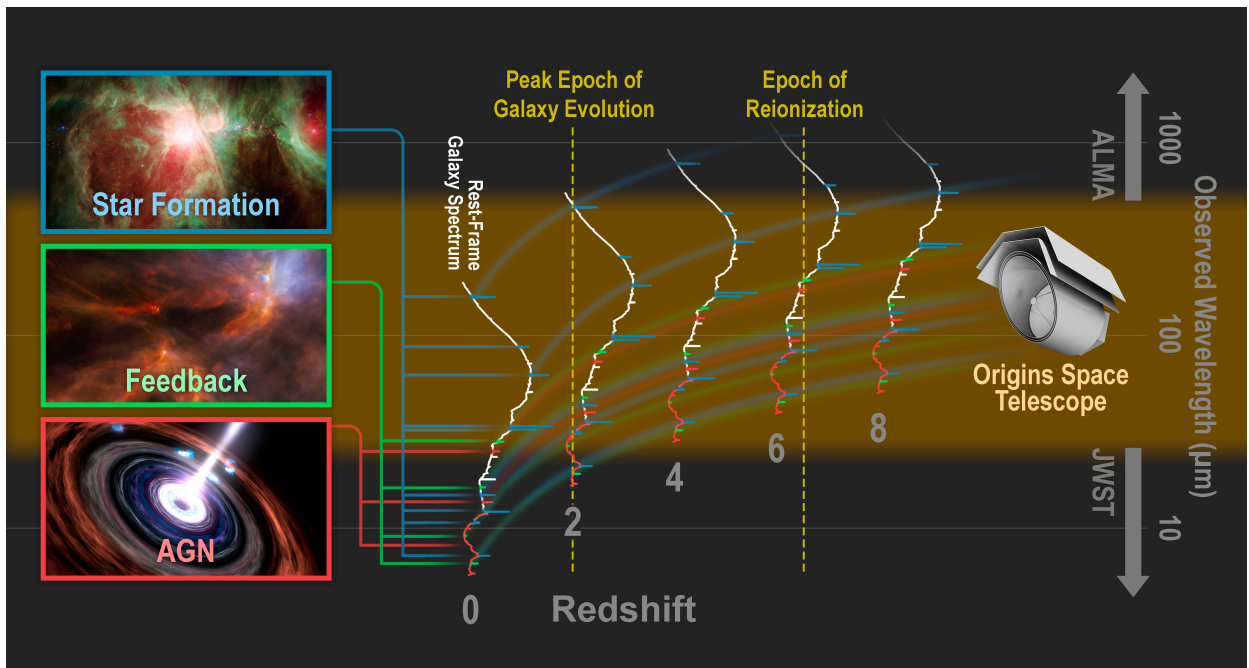

*Figure 2:* OST will simultaneously observe key diagnostics of star formation (e.g. PAH features and [NeII], [SiII], [OI], [CII]), feedback (e.g. $H_2$ lines), and black hole growth (e.g. [NeV], [OIV]) across millions of galaxies over cosmic time. Due to its ability to perform deep, wide spectroscopic surveys, OST will benefit from the factor of 1,000 increase in sensitivity while avoiding limitations imposed by source confusion.

Infrared observations provide unique tracers of the gas that feeds both star formation and black hole growth (see Figure 2) and is immune to the obscuring effects of dust that often hides young stars and galactic nuclei from view at shorter wavelengths (Spinoglio & Malkan 1992). The powerful energy emitted from growing supermassive black holes plays a critical role in regulating star formation in galaxies. OST will measure this feedback on the dense interstellar medium of nearby galaxies and will revolutionize our ability to identify and measure galactic outflows over the past 12 Gyr, linking theory and observations of how galaxies grow inside their dark matter halos.

A unique power of OST is its ability to conduct deep and wide spectroscopic surveys, taking full advantage of the large increase in sensitivity while avoiding many of the problems imposed by source crowding in photometric surveys. OST will conduct deep spectroscopic surveys over tens of square degrees, simultaneously measuring redshifts, star formation rates, and black hole accretion rates in millions of galaxies using bright, infrared emission lines from hot gas as well spectral features from dust grains excited by UV photons from young stars. These surveys will map out the cosmic history of star formation and black hole accretion over more than 95% of the age of the Universe. With the ability to trace the rise of metals, stars, and galaxies, and star formation and black hole growth from cosmic dawn to today, OST unveils the full story of our cosmic origins.

*The OST Mission Concepts*

OST is one of four Mission Concept studies sponsored by NASA for the 2020 NRC Decadal Study in Astronomy and Astrophysics. The OST study has two Mission Concepts, both covering wavelengths from 5–600 μm. Mission Concept 1 is comprised of a 9.1 m off-axis telescope and five instruments. Mission Concept 2, currently under development, is an optimized concept addressing all key science themes outlined in this article, composed of a 5.9 m on-axis telescope with JWST-sized collecting area and four instruments. The telescope in Mission Concept 2 is a three-mirror anastigmat, providing a field of view that allows OST to achieve all of its science goals. A field-steering fourth mirror enables rapid sky motion needed for the far-IR detectors. A multilayer sunshade, a cold baffle and cryocoolers ensure the telescope environment is maintained at 4 K, key to achieving OST's unprecedented sensitivity gain.

The four instruments under study for Mission Concept 2 enable a broad range of scientific advancements including those described here, but are designed to be flexible to the key questions of the 2030s and immense discovery space opened by OST. The **OST Survey Spectrometer (**OSS, resolution R~300 and R > 40,000 for 30-600 μm) will provide imaging and spectroscopy over large extragalactic fields and study protoplanetary disks. The **Mid-Infrared Imager, Spectrometer and Camera (MISC)** instrument (R~300) will conduct transit and emission spectroscopy of Jupiter to Earth-sized transiting exoplanets with simultaneous wavelength coverage from 5-25 μm, will have an imager and imaging spectrometer, and will also guide the observatory. OST's **Far-Infrared Imaging Polarimeter (FIP)** will conduct 40, 80, 120 and 240 μm imaging and polarimetry of wide extra-galactic fields and star-forming fields in our Galaxy. The **HEterodyne Receiver for OST (**HERO, resolution R~$10^6$-$10^7$ in selected wavelength bands between 111 μm and 566 μm) will measure the kinematics of gas in the interstellar medium and protoplanetary disks.

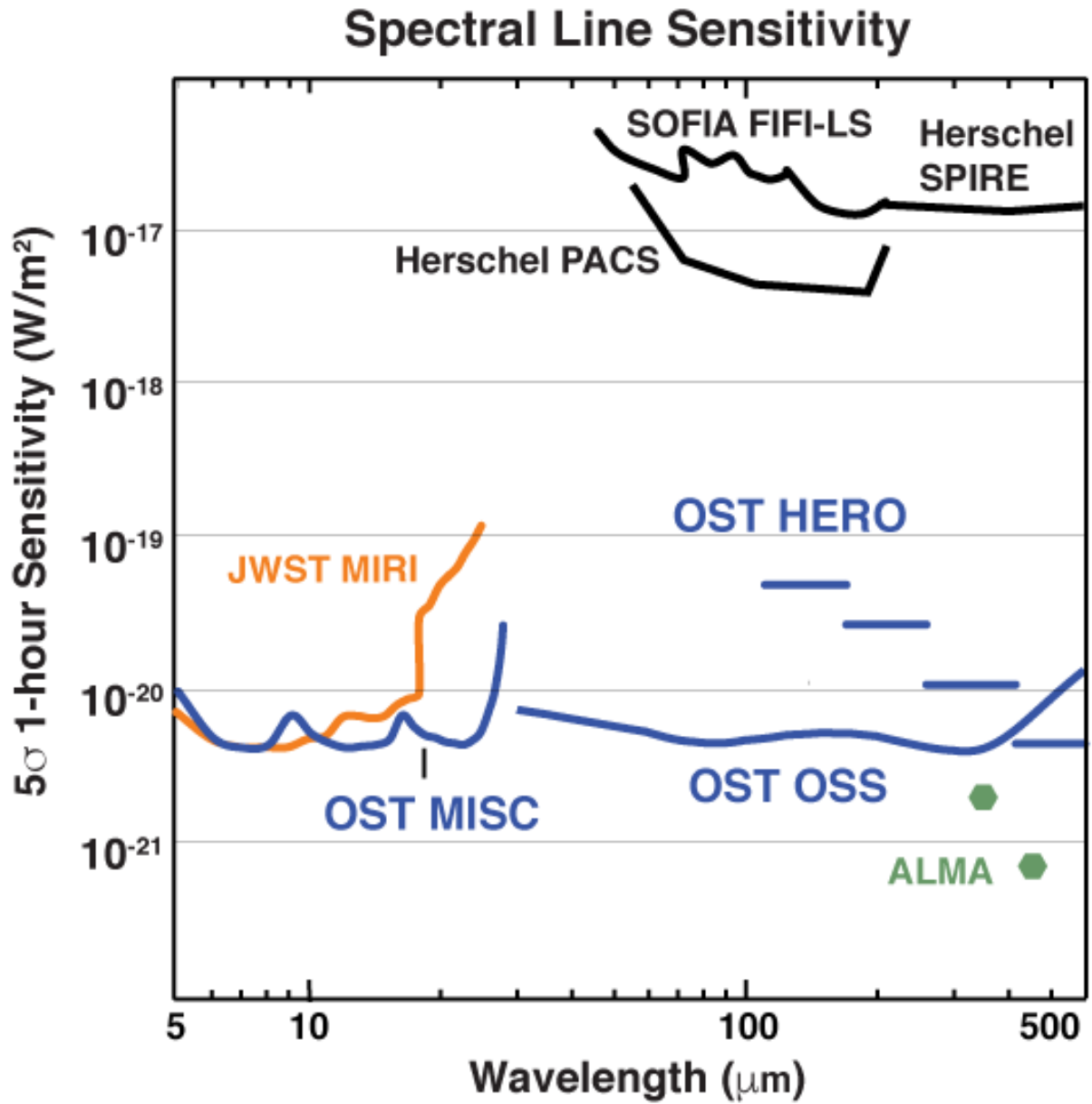

*Figure 3:* *OST offers expansive new discovery space, with an improvement in sensitivity of up to a factor of 1000 over any previous or planned facility. Spectral line sensitivity (5σ in 1 hour) across the far-IR band for the OST instruments (blue) for Concept 2 (5.9m) compared with existing and future facilities. The OSS and ALMA sensitivities refer to high-resolution ($R=10^6$) observations, for, e.g., protoplanetary disks and ISM kinematics.*

The observatory will be outfitted with large Control Moment Gyroscopes and momentum wheels to enable large-scale mapping of the sky at approximately 100 arcseconds per second. The OST Concept 2 design has minimal deployments (sunshields and solar arrays). The flight system, comprised of the telescope, instrument accommodation module, sunshield and spacecraft, will be launchable in a 7 or 8 m diameter fairing. Three possible launch vehicles – NASA's Space Launch System and two commercial vehicles – are expected to be equipped with such fairings by the 2030s, when OST is planned to launch.

*With a sensitivity gain of up to a factor of 1,000 over any previous or planned mission, OST will open unprecedented discovery space, allowing us to peer through an infrared window teeming with possibility. OST will fundamentally change our understanding of our cosmic origins — from the growth of galaxies and black holes, to uncovering the trail of water, to life signs in nearby Earth-size planets, and discoveries never imagined. Built to be highly adaptable, while addressing key science across many areas of astrophysics, OST will usher in a new era of infrared astronomy.*